# Controversy over Elastic Constants Based on Interatomic Potentials

L.G. Zhou and Hanchen Huang*

Department of Mechanical Engineering, University of Connecticut, Storrs, CT 06269

A controversy exists among literature reports of constraints on elastic constants. In particular, it has been reported that embedded atom method (EAM) potentials generally impose three constraints on elastic constants of crystals that are inconsistent with experiments. However, it can be shown that some EAM potentials do not impose such constraints at all. This paper first resolves this controversy by identifying the necessary condition when the constraints exist and demonstrating the condition is physically necessary. Furthermore, this paper reports that these three constraints are eliminated under all conditions, by using response EAM (R-EAM) potentials.

* Author to whom correspondence should be addressed; electronic mail: hanchen@uconn.edu

# 1 Introduction

Classical molecular mechanics simulations are common practice in materials and mechanics research, as highlighted in a recent review [1]. The widespread use of such simulations is partly due to the availability of more reliable interatomic potentials. For metals and alloys, the embedded atom method (EAM) potentials [2,3] are unambiguously more reliable than pair potentials in two ways. One, the Cauchy constraint $C_{12}=C_{44}$ is no longer imposed; here we use the Voigt notation in converting elastic constants $c_{\alpha\beta\mu\nu}$ to $C_{ij}$ [4]. Two, bond saturation and therefore surface relaxation are better represented. However, the EAM potentials appear to impose three other constraints on elastic constants and to be insufficient in describing the surface relaxation.

The first constraint that EAM potentials impose is $C_{12}>C_{44}$ according to the recent review [1]. As we show later, $C_{12}>C_{44}$ applies to only cubic crystals. This constraint is inconsistent with experimental measurements of certain cubic crystals such as body-centered cubic Cr [5], Ba [6], Cs [6], and Ge [7] (e.g. for Cr, $C_{12}=67.8$GPa and $C_{44}=100.8$GPa), in addition to Si as cited in the review. The second constraint that EAM potentials impose is $C_{13}>C_{44}$ for hexagonal-close-packed (HCP) crystals [8], it is in contradiction to experimental measurements of Be [9], Ru [6], and Y [10] (e.g. for Be, $C_{13}-C_{44}= -157$ GPa). The third constraint that EAM potentials impose is approximately $3C_{12}-C_{11}>2(C_{13}-C_{44})$ for HCP crystals [8], which is inconsistent with experimental measurements of Zn [11] and Cd [11] (e.g., for Zn, $3C_{12}-C_{11}=-65$ GPa and $2(C_{13}-C_{44})= 20$GPa). In contradiction to the claim of these three constraints on elastic constants [3,8], some EAM potentials do not impose such constraints. The EAM potential for Zn [12] gives $C_{11}=176$GPa, $C_{12}=50$GPa, $C_{13}=55$ GPa and $C_{44}=46$GPa; that is, it does not impose the constraint $3C_{12}-C_{11}>2(C_{13}-C_{44})$. And the potential for Be [12] gives $C_{13}=15$GPa, $C_{44}=168$GPa; that is, it does not impose the constraint $C_{13}>C_{44}$.

The first question is: do EAM potentials necessarily impose the three constraints on elastic constants? Then, if EAM potentials do impose these constraints that are inconsistent with experimental measurements, what solutions are available?

# 2 Elastic constants based on R-EAM and EAM potentials

To answer these two questions, we start with the response EAM (R-EAM) potential formulation [13] to derive the analytical expression of elastic constants $c_{\alpha\beta\mu\nu}$, and then achieve the corresponding expression based on EAM potentials as a special case. From the expression of $c_{\alpha\beta\mu\nu}$ based on EAM potentials, we identify the condition when EAM potentials impose these three constraints, and show that the elimination of these constraints within the EAM framework causes artifacts in the description of low coordination environments. From the expression of $c_{\alpha\beta\mu\nu}$ based on R-EAM potentials, we show that R-EAM potentials will not impose any of these three constraints.

Consider a perfect crystal subjected to deformation. For a given strain tensor $[\varepsilon]$, the displacement vector $\vec{r}_{ij}$ between atom $i$ and atom $j$ experiences a change $\Delta \vec{r}_{ij} = [\varepsilon]\vec{r}_{ij}$. After the change, the displacement vector becomes $\vec{r}_{ij}^{\,\varepsilon} = \vec{r}_{ij} + \Delta\vec{r}_{ij}$, and its magnitude becomes $r_{ij}^{\varepsilon} = \sqrt{r_{ij}^2 + \Delta r_{ij}^2 + 2\vec{r}_{ij}\cdot\Delta\vec{r}_{ij}}$. The magnitude of this change $\Delta r_{ij}$ is therefore $\Delta r_{ij} = r_{ij}^{\varepsilon} - r_{ij} = \Delta\vec{r}_{ij}\cdot\vec{r}_{ij}/r_{ij} - \frac{1}{2r_{ij}}\left(\Delta\vec{r}_{ij}\cdot\vec{r}_{ij}/r_{ij}\right)^2 + \frac{1}{2r_{ij}}\Delta\vec{r}_{ij}\cdot\Delta\vec{r}_{ij}$. For a perfect crystal, we can focus on one atom $i$ since all atoms are identical. The implicit assumption is that all atoms remain identical even after deformation. This is approximation when relative relaxation happens such as in HCP metals; the relative relaxation will only affect $C_{11}$-$C_{12}$ [8]. With the focus on atom $i$, we drop the subscript $i$ for simplicity to write:

$$\Delta r_{im} \equiv \Delta r_m = \frac{1}{r_m}\sum_{\alpha,\beta=1,3}\varepsilon^{\alpha\beta}r_m^{\alpha}r_m^{\beta} - \frac{1}{2r_m}\left(\frac{1}{r_m}\sum_{\alpha,\beta=1,3}\varepsilon^{\alpha\beta}r_m^{\alpha}r_m^{\beta}\right)^2 + \frac{1}{2r_m}\sum_{\alpha=1,3}\left(\sum_{\beta=1,3}\varepsilon^{\alpha\beta}r_m^{\beta}\right)^2. \tag{1}$$

According to R-EAM potentials [13], the potential energy of atom $i$ in a perfect crystal (consisting of equivalent atoms) is:

$$E = F_m + \Phi_m = F\left(\sum_m \rho(r_m)\right) + \frac{1}{2}\sum_m \phi(r_m)\left[1 + R\left(\sqrt{\sum_{l\neq m}\rho(r_l)\sum_{m\neq l}\rho(r_m)}\right)\right]$$

$$= F\left(\sum_m \rho(r_m)\right) + \frac{1}{2}\sum_m \phi(r_m)\left[1 + R\left(\sum_{l\neq m}\rho(r_l)\right)\right]. \tag{2}$$

The first term represents the embedding energy $F_m$ and the second term the pair-interaction energy $\Phi_m$; where $F$ is the embedding function, $\phi$ the pair-interaction function, $\rho$ the electron density contribution from an atom, and $R$ the response function. Under the strain $[\varepsilon]$, the embedding energy and the pair-interaction energy become:

$$F_m^{\varepsilon} = F\left(\sum_m \rho(r_m + \Delta r_m)\right) \text{ and } \Phi_m^{\varepsilon} = \frac{1}{2}\sum_m \phi(r_m + \Delta r_m)\left[1 + R\left(\sum_{l\neq m}\rho(r_l + \Delta r_l)\right)\right]. \tag{3}$$

The strain energy associated with the atom $i$ is:
$$\Delta E = F_m^{\varepsilon} + \Phi_m^{\varepsilon} - F_m - \Phi_m. \tag{4}$$

For a given atomic volume $\Omega$ of the crystal, the strain energy density is $\Delta E/\Omega$, and the elastic constants $c_{\alpha\beta\mu\nu}$ are:

$$c_{\alpha\beta\mu\nu} = \frac{\partial^2(\Delta E/\Omega)}{\partial\varepsilon_{\alpha\beta}\partial\varepsilon_{\mu\nu}} = \sum_{i=1}^{6}t_i \tag{5}$$

Based on the derivation in the Appendix, the seven terms in equation (5) are:

$$t_1 = \frac{1}{2\Omega}\sum_m \phi(r_m)R'(\bar{\rho}_{0m})\left[\sum_{l\neq m}(\rho''(r_l) - \rho'(r_l)/r_l)\frac{r_l^{\alpha}r_l^{\beta}r_l^{\mu}r_l^{\nu}}{r_l^2}\right] \tag{5a}$$

$$t_2 = \frac{1}{2\Omega}\sum_m \phi(r_m) R''(\bar{\rho}_{0m})\left(\sum_{l\neq m}\rho'(r_l)\frac{r_l^\alpha r_l^\beta}{r_l}\right)\left(\sum_{l\neq m}\rho'(r_l)\frac{r_l^\mu r_l^\nu}{r_l}\right) \quad (5b)$$

$$t_3 = \frac{1}{2\Omega}\sum_m \phi'(r_m) R'(\bar{\rho}_{0m})\left[\sum_{l\neq m}\rho'(r_l)\left(\frac{r_l^\alpha r_l^\beta}{r_l}\frac{r_m^\mu r_m^\nu}{r_m} + \frac{r_l^\mu r_l^\nu}{r_l}\frac{r_m^\alpha r_m^\beta}{r_m}\right)\right] \quad (5c)$$

$$t_4 = \frac{1}{2\Omega}\left[1+R(\bar{\rho}_{0m})\right]\left[\sum_m \left(\phi''(r_m) - \phi'(r_m)/r_m\right)\frac{r_m^\alpha r_m^\beta r_m^\mu r_m^\nu}{r_m^2}\right] \quad (5d)$$

$$t_5 = \frac{F'(\bar{\rho}_0)}{\Omega}\sum_m \left(\rho''(r_m) - \rho'(r_m)/r_m\right)\left(\frac{r_m^\alpha r_m^\beta r_m^\mu r_m^\nu}{r_m^2}\right) \quad (5e)$$

$$t_6 = \frac{F''(\bar{\rho}_0)}{\Omega}\left(\sum_m \rho'(r^m)\frac{r_m^\alpha r_m^\beta}{r_m}\right)\left(\sum_m \rho'(r^m)\frac{r_m^\mu r_m^\nu}{r_m}\right). \quad (5f)$$

Here, a single prime indicates a first-order derivative, a double prime indicates a second-order derivative, the summation over $m$ runs over all interacting neighbors of atom $i$, and $\bar{\rho}_0 = \sum_m \rho(r_m)$ and $\bar{\rho}_{0m} = \bar{\rho}_0 - \rho(r_m)$ are values at equilibrium under no strain. Taking the response function $R$ as constant zero, we recover the expression $c_{\alpha\beta\mu\nu}$ based on EAM potentials as:

$$c_{\alpha\beta\mu\nu} = p_1 + p_2 + p_3 \quad (6)$$

where

$$p_1 = \frac{1}{2\Omega}\sum_m \left(\phi''(r_m) - \phi'(r_m)/r_m\right)\left(\frac{r_m^\alpha r_m^\beta r_m^\mu r_m^\nu}{r_m^2}\right) \quad (6a)$$

$$p_2 = \frac{F'(\bar{\rho}_0)}{\Omega}\sum_m \left(\rho''(r_m) - \rho'(r_m)/r_m\right)\left(\frac{r_m^\alpha r_m^\beta r_m^\mu r_m^\nu}{r_m^2}\right) \quad (6b)$$

$$p_3 = \frac{F''(\bar{\rho}_0)}{\Omega}\left(\sum_m \rho'(r^m)\frac{r_m^\alpha r_m^\beta}{r_m}\right)\left(\sum_m \rho'(r^m)\frac{r_m^\mu r_m^\nu}{r_m}\right). \quad (6c)$$

## 3  Resolution to the controversy over constraints on elastic constants

With equations (5) and (6), we now answer the two questions posed at the start of this paper. Let us examine the first constraint according to EAM potentials. From equation (6) and the symmetry of cubic crystals $\sum_m x_m^2 y_m^2 = \sum_m y_m^2 z_m^2$, we have:

$$(C_{12} - C_{44}) \equiv (c_{1122} - c_{2323}) = \frac{F''(\bar{\rho}_0)}{\Omega}\left(\sum_m \rho'(r^m)\frac{x_m^2}{r_m}\right)^2. \quad (7)$$

Given that $F''(\bar{\rho}_0)$ is positively defined[1], the $(C_{12} - C_{44})$ term is always positive for cubic crystals, unless the terms in the sum precisely cancel out each other. That is, the first constraint $C_{12} > C_{44}$ always exists based on EAM potentials. Examining the second constraint, we note that the symmetry of HCP crystals gives $\sum_m x_m^2 z_m^2 = \sum_m y_m^2 z_m^2$. Consequently,

$$(C_{13} - C_{44}) \equiv (c_{1133} - c_{2323}) = \frac{F''(\bar{\rho}_0)}{\Omega} \left( \sum_m \rho'(r^m) \frac{x_m^2}{r_m} \right) \left( \sum_m \rho'(r^m) \frac{z_m^2}{r_m} \right). \qquad (8)$$

Cubic symmetry will also result in equation (8), which then reduces to equation (7) under such symmetry. This quantity $(C_{13} - C_{44})$ is always positive so the second constraint $C_{13} > C_{44}$ exists for HCP crystals, provided that the electron density $\rho(r)$ is a monotonic function of distance $r$. This condition is indeed met by most EAM potentials, but *not all*. Examining the third constraint, we note that the symmetry of HCP crystals gives $\sum_m x_m^4 = \sum_m y_m^4 = 3 \sum_m x_m^2 y_m^2$. Consequently:

$$(C_{11} - C_{12}) \equiv (c_{1111} - c_{1122})$$
$$= \frac{1}{2\Omega} \sum_m \left( \phi''(r_m) - \phi'(r_m)/r_m \right) \left( \frac{2 x_m^2 y_m^2}{r_m^2} \right)$$
$$+ \frac{F'(\bar{\rho}_0)}{\Omega} \sum_m \left( \rho''(r_m) - \rho'(r_m)/r_m \right) \left( \frac{2 x_m^2 y_m^2}{r_m^2} \right) - W. \qquad (9)$$

Here $W$ is the contribution of sub-lattice relaxation on elastic constant, and is positive [8]. The sub-lattice relaxation has a contribution to elastic constants in HCP crystals only to $(C_{11} - C_{12})$, and has no impacts on $C_{11}+C_{12}$, $C_{13}$ and $C_{44}$. Using the same symmetry consideration for HCP crystals $\sum_m x_m^4 = \sum_m y_m^4 = 3 \sum_m x_m^2 y_m^2$, we also have:

$$(C_{11} + C_{12}) \equiv (c_{1111} + c_{1122})$$
$$= \frac{1}{2\Omega} \sum_m \left( \phi''(r_m) - \phi'(r_m)/r_m \right) \left( \frac{4 x_m^2 y_m^2}{r_m^2} \right) / r_m^2$$
$$+ \frac{F'(\bar{\rho}_0)}{\Omega} \sum_m \left( \rho''(r_m) - \rho'(r_m)/r_m \right) \left( \frac{4 x_m^2 y_m^2}{r_m^2} \right) / r_m^2 + \frac{2 F''(\bar{\rho}_0)}{\Omega} \left( \sum_m \rho'(r^m) \frac{x_m^2}{r_m} \right)^2. \qquad (10)$$

Combining equations (9) and (10), we have

$$(3 C_{12} - C_{11}) = (C_{11} + C_{12}) - 2(C_{11} - C_{12}) = \frac{2 F''(\bar{\rho}_0)}{\Omega} \left( \sum_m \rho'(r^m) \frac{x_m^2}{r_m} \right)^2 + 2W. \qquad (11)$$

Comparing equations (11) with equation (8), we have:

$$(3 C_{12} - C_{11}) - a(C_{13} - C_{44}) = 2W \text{ with } a = 2 \left( \sum_m \rho'(r^m) \frac{x_m^2}{r_m} \right) / \left( \sum_m \rho'(r^m) \frac{z_m^2}{r_m} \right). \qquad (12)$$

For HCP crystals, $a$ is about 2. Since $W$ is positive, the third constraint $(3C_{12}-C_{11})>2(C_{13}-C_{44})$ exists for HCP crystals, provided that $\rho(r)$ is a monotonic function of distance $r$.

The analyses of equation (6) based on EAM potentials show that (1) the first constraint exists for cubic crystals, (2) the second constraint exists for HCP crystals provided that $\rho(r)$ is a monotonic function of $r$, and (3) the third constraint exists approximately (with $a$ being approximately (2) for HCP crystals provided that $\rho(r)$ is a monotonic function of $r$.

The key in the last two constraints is the condition that $\rho(r)$ is a monotonic function of $r$. The last two constraints – which are particularly relevant to HCP crystals – do not necessarily exist if $\rho(r)$ is *not* monotonic. Indeed, the EAM potentials for Be and Zn [12] use non-monotonic $\rho(r)$. As a result, these potentials do not impose any of the two constraints mentioned above for HCP crystals. However, the embedding energy will be non-monotonic as a function of distance, because of $\rho(r)$ being non-monotonic. Such embedding energy function may result in unphysical binding states of low coordination atoms. To show this unphysical nature, we consider dimers, trimers, and tetramers of Zn, and compare EAM results with results from available R-EAM based calculations and quantum mechanics calculations as described in [13]; each cluster is always equilateral. As shown in Figure 1(a), the EAM potential with non-monotonic $\rho(r)$ indeed gives rise to binding states of trimer and tetramer that are inconsistent with quantum mechanics calculations.

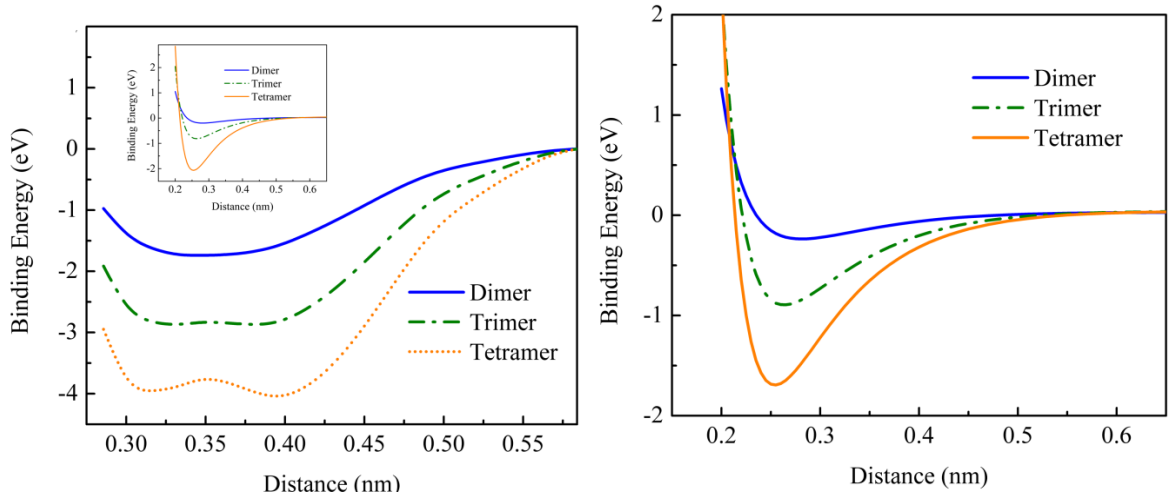

Fig. 1 Binding energies of a dimer, a trimer, and a tetramer of Zn as a function of the distance between any two atoms; according to (a) the EAM potential [12] and (b) the R-EAM potential. The inset shows the quantum mechanics results.

According to the analytical expression of equation (6) and the numerical results of Figure 1(a), EAM potentials will necessarily impose the three constraints on elastic constants, without causing unphysical binding of low coordination atoms. The R-EAM potential, in contrast, provides a description of binding among low coordination atoms that is consistent with quantum mechanics results; Figure 1(b). More importantly, R-EAM potentials do not impose any of the three constraints on elastic constants, as explained below.

For the first constraint, we use the formulation of $c_{\alpha\beta\mu\nu}$ based on R-EAM potentials (and based on cubic symmetry), yielding:

$$(C_{12} - C_{44}) = \frac{F''(\bar{\rho}_0)}{\Omega}\left(\sum_m \rho'(r^m) \frac{x_m^2}{r_m}\right)^2$$
$$+ \frac{1}{2\Omega}\sum_m \phi(r_m) R''(\bar{\rho}_{0m})\left(\sum_{l \neq m} \rho'(r_l) \frac{x_l^2}{r_l}\right)^2$$
$$+ \frac{1}{\Omega}\sum_m \phi'(r_m) R'(\bar{\rho}_{0m}) \left(\frac{y_m^2}{r_m}\right)\left(\sum_{l \neq m} \rho'(r_l) \frac{x_l^2}{r_l}\right). \tag{13}$$

The first term is the same as that in equation (7) and is always positive. However, a given function and its derivatives can independently be positive or negative. For example, even if the pair interaction term $\phi(r)$ is chosen as always positive, its derivative $\phi'(r)$ does not need to be and cannot be always positive. In addition, the first and second derivatives of the response function each can independently be either positive or negative. That is, the last two terms of equation can be either positive or negative, so there is no constraint $(C_{12} - C_{44}) > 0$ according to R-EAM potentials.

For the second constraint on elastic constants based on R-EAM potentials, we consider HCP symmetry, giving:

$$(C_{13} - C_{44}) = \frac{F''(\bar{\rho}_0)}{\Omega}\left(\sum_m \rho'(r^m) \frac{x_m^2}{r_m}\right)\left(\sum_m \rho'(r^m) \frac{z_m^2}{r_m}\right)$$
$$+ \frac{1}{2\Omega}\sum_m \phi(r_m) R''(\bar{\rho}_{0m})\left(\sum_{l \neq m} \rho'(r_l) \frac{x_l^2}{r_l}\right)\left(\sum_{l \neq m} \rho'(r_l) \frac{z_l^2}{r_l}\right)$$
$$+ \frac{1}{2\Omega}\sum_m \frac{\phi'(r_m) R'(\bar{\rho}_{0m})}{r_m}\left[\sum_{l \neq m} \frac{\rho'(r_l)}{r_l}\left(x_l^2 z_m^2 + x_m^2 z_l^2\right)\right]. \tag{14}$$

The first term is the same as that in equation (8) and is always positive provided that $\rho(r)$ is a monotonic function. Following the same logic as for equation (13), the last two terms of equation (14) each can independently be either positive or negative, so there is no constraint $(C_{13} - C_{44}) > 0$ according to R-EAM potentials.

For the third constraint on elastic constants based on R-EAM potentials, we consider HCP symmetry,

$$(3C_{12} - C_{11}) - a(C_{13} - C_{44})$$
$$= 2W'$$

$$+\frac{1}{\Omega}\sum_{m}\left[\phi(r_m)R''(\bar{\rho}_{0m})\left(\sum_{l\neq m}\rho'(r_l)\frac{x_l^2}{r_l}\right)^2\left(1-a\frac{\sum_{l\neq m}\rho'(r_l)\frac{z_l^2}{r_l}}{\sum_{l\neq m}\rho'(r_l)\frac{x_l^2}{r_l}}\right)\right]$$

$$+\frac{1}{\Omega}\sum_{m}\left[\frac{\phi'(r_m)R'(\bar{\rho}_{0m})}{r_m}\left(\sum_{l\neq m}\frac{\rho'(r_l)}{r_l}\left(2x_m^2 x_l^2 - az_m^2 x_l^2 - ax_m^2 z_l^2\right)\right)\right]. \tag{15}$$

The first term ($2W'$), based on the same argument as that in equation (12), is always positive [8]. Following the same logic as for equation (13), the last two terms of equation (15) each can independently be either positive or negative, so there is no constraint $(3C_{12} - C_{11}) > 2(C_{13} - C_{44})$ according to R-EAM potentials; the same approximation $a \approx 2$ is used.

## 4 Conclusions

In summary, this paper first resolves a controversy on whether EAM potentials impose three constraints on elastic constants. Using analytical formulations, we have demonstrated that the EAM potentials always impose the first constraint $C_{12} > C_{44}$ for cubic crystals, that they impose the second constraint $C_{13} > C_{44}$ and approximately the third constraint $(3C_{12} - C_{11}) > 2(C_{13} - C_{44})$ for HCP crystals provided that electron density $\rho(r)$ is a monotonic function of $r$. When an EAM potential uses non-monotonic r, the last two constraints do not necessarily exist, but the potential provides an unphysical description of binding among low coordination atoms. In addition to resolving the controversy, we have reported the analytical expression of elastic constants based on R-EAM potentials and shown that R-EAM potentials do not impose any of these three constraints on elastic constants.

**Acknowledgement:** The authors gratefully acknowledge financial support from the Department of Energy Office of Basic Energy Sciences (DE-FG02-09ER46562).

# Appendix: Derivation of Elastic Constants $c_{\alpha\beta\mu\nu}$

Here, we present details of derivation of the elastic constants in equation (5). Starting from equation (4), and keeping terms up to the second order of displacement $\Delta r$, the energy change is:

$$\Delta E = \frac{1}{2}\sum_m \phi(r_m)\left[ R'\left(\sum_{l\neq m}\rho(r_l)\right)\left(\sum_{l\neq m}\rho'(r_l)\Delta r_l + \frac{1}{2}\sum_{l\neq m}\rho''(r_l)(\Delta r_l)^2\right) + \frac{1}{2}R''\left(\sum_{l\neq m}\rho(r_l)\right)\left(\sum_{l\neq m}\rho'(r_l)\Delta r_l\right)^2 \right]$$

$$+ \frac{1}{2}\sum_m \phi'(r_m)\Delta r_m \left[ 1 + R\left(\sum_{l\neq m}\rho(r_l)\right) + R'\left(\sum_{l\neq m}\rho(r_l)\right)\left(\sum_{l\neq m}\rho'(r_l)\Delta r_l\right) \right]$$

$$+ \frac{1}{4}\sum_m \phi''(r_m)(\Delta r_m)^2 \left[ 1 + R\left(\sum_{l\neq m}\rho(r_l)\right) \right]$$

$$+ F'(\bar{\rho}_0)\sum_m \rho'(r_m)\Delta r_m + \frac{1}{2}F'(\bar{\rho}_0)\sum_m \rho''(r_m)(\Delta r_m)^2 + \frac{1}{2}F''(\bar{\rho}_0)\left(\sum_m \rho'(r^m)\Delta r_m\right)^2 \quad (A1)$$

Using equation (1) in the paper, keeping terms up to the second order of strain in equation (A1), and combining with the inversion symmetry of a crystal, we have:

$$\Delta E = \sum_{i=1}^{6} a_i \quad (A2)$$

$$a_1 = \frac{1}{4}\sum_m \phi(r_m)R'\left(\sum_{l\neq m}\rho(r_l)\right)\left[\sum_{l\neq m}\rho''(r_l)\left(\frac{1}{r_l}\sum_{\alpha,\beta=1,3}\varepsilon^{\alpha\beta}r_l^\alpha r_l^\beta\right)^2\right]$$

$$+ \frac{1}{2}\sum_m \phi(r_m)R'\left(\sum_{l\neq m}\rho(r_l)\right)\left[\sum_{l\neq m}\rho'(r_l)\left(\frac{1}{r_l}\sum_{\alpha,\beta=1,3}\varepsilon^{\alpha\beta}r_l^\alpha r_l^\beta - \frac{1}{2r_l}\left(\frac{1}{r_l}\sum_{\alpha,\beta=1,3}\varepsilon^{\alpha\beta}r_l^\alpha r_l^\beta\right)^2 + \frac{1}{2r_l}\left(\sum_{\alpha,\beta=1,3}\varepsilon^{\alpha\beta}\varepsilon^{\alpha\beta}r_l^\alpha r_l^\alpha\right)\right)\right]$$

$$a_2 = \frac{1}{4}\sum_m \phi(r_m)R''\left(\sum_{l\neq m}\rho(r_l)\right)\left[\left(\sum_{l\neq m}\rho'(r_l)\frac{1}{r_l}\sum_{\alpha,\beta=1,3}\varepsilon^{\alpha\beta}r_l^\alpha r_l^\beta\right)^2\right]$$

$$a_3 = \frac{1}{2}\sum_m \phi'(r_m)R'\left(\sum_{l\neq m}\rho(r_l)\right)\left(\frac{1}{r_m}\sum_{\alpha,\beta=1,3}\varepsilon^{\alpha\beta}r_m^\alpha r_m^\beta\right)\left[\sum_{l\neq m}\rho'(r_l)\left(\frac{1}{r_l}\sum_{\alpha,\beta=1,3}\varepsilon^{\alpha\beta}r_l^\alpha r_l^\beta\right)\right]$$

$$a_4 = \frac{1}{4}\sum_m \phi''(r_m)\left[1 + R\left(\sum_{l\neq m}\rho(r_l)\right)\right]\left(\frac{1}{r_m}\sum_{\alpha,\beta=1,3}\varepsilon^{\alpha\beta}r_m^\alpha r_m^\beta\right)^2$$

$$+ \frac{1}{2}\sum_m \phi'(r_m)\left[1 + R\left(\sum_{l\neq m}\rho(r_l)\right)\right]\left[\frac{1}{r_m}\sum_{\alpha,\beta=1,3}\varepsilon^{\alpha\beta}r_m^\alpha r_m^\beta - \frac{1}{2r_m}\left(\frac{1}{r_m}\sum_{\alpha,\beta=1,3}\varepsilon^{\alpha\beta}r_m^\alpha r_m^\beta\right)^2 + \frac{1}{2r_m}\left(\sum_{\alpha,\beta=1,3}\varepsilon^{\alpha\beta}\varepsilon^{\alpha\beta}r_m^\alpha r_m^\alpha\right)\right]$$

$$a_5 = \frac{1}{2}F'(\bar{\rho}_0)\sum_j \rho''(r_m)\left(\frac{1}{r_m}\sum_{\alpha,\beta=1,3}\varepsilon^{\alpha\beta}r_m^\alpha r_m^\beta\right)^2$$

$$+ F'(\bar{\rho}_0)\sum_m \rho'(r_m)\left[\frac{1}{r_m}\sum_{\alpha,\beta=1,3}\varepsilon^{\alpha\beta}r_m^\alpha r_m^\beta - \frac{1}{2r_m}\left(\frac{1}{r_m}\sum_{\alpha,\beta=1,3}\varepsilon^{\alpha\beta}r_m^\alpha r_m^\beta\right)^2 + \frac{1}{2r_m}\left(\sum_{\alpha,\beta=1,3}\varepsilon^{\alpha\beta}\varepsilon^{\alpha\beta}r_m^\alpha r_m^\alpha\right)\right]$$

$$a_6 = \frac{1}{2} F''(\bar{\rho}_0) \left[ \sum_m \rho'(r^m) \left( \frac{1}{r_m} \sum_{\alpha,\beta=1,3} \varepsilon^{\alpha\beta} r_m^\alpha r_m^\beta \right) \right]^2$$

Taking the second order derivative of equation (A2), one obtains $\dfrac{\partial^2 (\Delta E / \Omega)}{\partial \varepsilon_{\alpha\beta} \partial \varepsilon_{\mu\nu}}$. The corresponding $a_i$ in equation (A2) will then become $t_i$ in equation (5).